\documentclass[12pt]{article}
\usepackage[dvips]{graphicx}
\usepackage{amssymb}
\usepackage{amsmath}

\def\beq{\begin{equation}}\def\eeq{\end{equation}}
\def\bea{\begin{eqnarray}}\def\eea{\end{eqnarray}}

\begin{document}

\title{Restoring locality of scalar fields on a causal set by avoiding the use of d'Alembertians}
\author{Roman Sverdlov,
\\Department of Mathematics, University of New Mexico} 

\date{June 26, 2018}
\maketitle

\begin{abstract}
In this paper we address the non-locality issue of quantum field theory on a causal set by rewriting it in such a way that avoids the use of d'Alembertian. We do that by replacing scalar field over points with scalar field over edges, where the edges are taken to be very long rather than very short. In particular, they are much longer than the size of the laboratory. Due to their large length, we can single out the edges that are almost parallel to each other, and then use directional derivatives in the direction of those edges (as opposed to d'Alembertian) along with a constraint that the derivatives are small in the direction perpendicular to those edges, in order to come up with a plane wave. The scalar field is thought to reside at the future end of those edges, which renders the seemingly nonlocal effects of their large length as physically irrelevant. After that we add by hand the interaction of those plane waves that would amount to 4-vertex coupling of plane waves. 
\end{abstract}

\subsection*{Introduction}

A causal set is a model of a spacetime as a partially ordered set, where partial ordering, $\prec$, corresponds to lightcone causal relations: $a \prec b$ if and only if point $a$ is before $b$ in Lorentzian sense (for more detailed review, see \cite{Review1} and \cite{Review2}). It has been shown (see \cite{Hawking} and \cite{Malament}) that there is a one to one correspondence between $\prec$ and $g_{\mu \nu} / \vert g \vert$; in discrete case, $\vert g \vert$ can be deduced from the count of points, under the assumption that every point takes up the same volume, which means that $\prec$, indeed, provides the complete description of discretized geometry. Logically, this implies that any given Lorentz covariant expression could be rewritten without the use of Lorentzian indexes. In particular, the Lorentz contraction between the two derivatives should be rewritten in terms of some type of sum, without derivative signs; vector fields can be replaced with scalar two-point functions, and so forth. There have, in fact, been proposals on defining d'Alembertians in terms of causal relations alone (see \cite{Dambertian1} and \cite{Dambertian2}) but they are non-local due to the fact that the Lorentzian $\epsilon$-neighborhood of a point fills in the vicinity of its light cone and has infinite volume (for more details of discussion of non-locality issue see \cite{Nonlocality} and \cite{NonlocalityTheorem}). 

However, unlike d'Alembertian, the directional derivative can, in fact, be modeled locally. Thus, we propose to tackle the nonlocality issue by replacing the continuum theory involving d'Alembertian with the one involving directional derivatives, and then discretizing the latter. In particular, for the $\phi^4$-theory,
\beq {\cal L} (\phi;x)  = - \frac{1}{2} \phi \Delta \phi - \frac{m^2}{2} \phi^2 - \frac{\lambda}{4!} \phi^4 \label{Conventional} \eeq
we can replace $\phi (x)$ with $\phi (x,v)$, where $v$ is a timelike unit tangent vector ($v^{\mu} v_{\mu}=1$), replace ${\cal L} (\phi; x)$ given above with ${\cal L} (\phi; x, v)$ given as 
\beq {\cal L} (\phi; x, v) = \frac{1}{2} v^{\mu} v^{\nu} \frac{\partial \phi}{\partial x^{\mu}} \frac{\partial \phi}{\partial x^{\nu}} + \frac{m^2}{2} \phi^2 + \nonumber \eeq
\beq + \int dv_1 dv_2 dv_3 \frac{\lambda (v, v_1, v_2, v_3)}{4!} \phi (x, v) \phi (x,v_1) \phi (x,v_2) \phi (x,v_3) \label{Unconventional} \eeq
where $\omega (v, v_1, v_2, v_3) \rightarrow 0$ in the limit of $v_{1,2,3} \rightarrow \infty$, and introduce a constraint 
\beq (g^{\mu \nu} - v^{\mu} v^{\nu} ) \frac{\partial^2 \phi}{\partial x^{\mu} \partial x^{\nu}}   = - \epsilon (R(x,v)) \phi (x,v) \label{Constraint} \eeq
where $R(x,v)$ is a curvature, which would pick up $v$-dependence in discrete situation, and $\epsilon (R)$ is a positive-valued function whose value is typically very small unless $R$ is extremely large; we may assume $\epsilon (0) =0$. The reason $R$ is $v$-dependent is that arbitrarily large Lorentz boost of $R_{x^{\mu} x^{\mu}}$ would have very large components unless something is done to prevent it. Thus, we modify the theory and claim that $R$ has $v$-dependence that is only felt if $v$ is close to $c$. Thus, as long as $v$ is far away from $c$, we assume that $v$-dependence of $R$ is negligible, and, therefore, it transforms under Lorentz transformations as usual. But, once the value of $v$ becomes very large, the effects of $v$-dependence would become large as well, and they would prevent the components of $R_{x^{\mu} x^{\nu}}$ from becoming large. Said $v$-dependence wouldn't violate Lorentz invariance since we are already used to the fact that $x$-dependence doesn't violate translational invariance, and $v$ is just another coordinate if we replace a manifold with a tangent bundle -- which is ultimately what we are doing. In any case, attempting to actually define discretized version of curvature is beyond the scope of this paper; we will simply develop a Lagrangian of scalar fields under the assumption that there is some agreed-upon definition of curvature (a couple of rather different examples of such is given in \cite{Curvature} and \cite{GravityAndMatter}). To see what we just have done, consider the frame where $v^{\mu}= \delta^{\mu}_0$: 
\beq v^{\mu} = \delta^{\mu}_0 \Longrightarrow \frac{1}{2} v^{\mu} v^{\nu} \frac{\partial \phi}{\partial x^{\mu}} \frac{\partial \phi}{\partial x^{\nu}} = \frac{1}{2} \bigg( \frac{\partial \phi}{\partial x^0} \bigg)^2 \eeq
\beq v^{\mu} = \delta^{\mu}_0 \Longrightarrow (g^{\mu \nu} - v^{\mu} v^{\nu}) \frac{\partial^2 \phi}{\partial x^{\mu} \partial x^{\nu}} = - \frac{\partial^2 \phi}{\partial x^k \partial x^k} \eeq
If we assume spacetime is flat (and, accordingly, $\epsilon (R) = \epsilon (0) =0$) and if we also assume $\lambda (v,v_1,v_2,v_3)= 0$, this will result in plane waves, which is exactly what we would get if we assume $\lambda=0$ in the conventional Lagrangian, Eq \ref{Conventional}, again in flat spacetime. Due to the fact that $v$ is now viewed as a separate variable, the interaction term is no longer local, which is the reason for an integral sign. But, due to the fact that $\lambda (v, v_1, v_2, v_3) \rightarrow 0$ as $v_{1,2,3} \rightarrow \infty$, the divergences are avoided; roughly speaking, $w(v,v_1,v_2,v_3)$ plays the role of the ultraviolet cutoff. There is no need to impose a conservation of $v$ constraint on that integral because the harmonics whose total momentum, as defined through wave-vector, isn't conserved will integrate to zero, and the $v$-variable approximates wave-vector rather closely due to the constraint given in Eq \ref{Constraint}. 

The point of doing all this is, like we said earlier, to avoid d'Alembertian, hereby regaining locality. In other words, we claim that the terms in Eq \ref{Unconventional} do, in fact, have local representation in a causal set, in contrast to Eq \ref{Conventional}. The subject of the rest of the paper is to do the causal set discretization of Eq \ref{Unconventional} and \ref{Constraint}. 

\subsection*{Causal set discretization}

Causal set is a partially ordered set, in which partial ordering, $\prec$, is interpreted as lightcone causal relation. It is assumed to be discrete in a sense that for any points $p$ and $q$ the number of points $r$ satisfying $p \prec r \prec q$ is at most finite. The past and future lightcones of $p$ are denoted by $J^- (p)$ and $J^+ (p)$, respectively: 
\beq J^- (p) = \{ r \vert r \prec p \} \eeq
\beq J^+ (p) = \{ r \vert r \succ p \} \eeq
We also introduce direct neighbor relation, $\prec^*$, as follows: $p \prec^* q$ if and only if $p \prec q$ holds and there is no point $r$ satisfying $p \prec r \prec q$. The distance between the points $p \prec q$ is given by (see \cite{Distance})
\beq \tau (p,q) = \xi \max \{n \vert \exists r_1, \ldots, r_{n-1} (p \prec r_1 \prec \ldots \prec r_{n-1} \prec q) \} \label{Length} \eeq
where $\xi$ is some small number (typically assumed to be Plank's scale) and the use of $\max$ instead of $\min$ is due to the minus signs in Minkowskian metric (our convention is $(+ - - -)$). The longest chain of points in Eq \ref{Length} is a discretized version of a \emph{geodesics} connecting points $p$ and $q$. 

The discretized version of $(x,v)$ is $(p,q)$, where $q$ coincides with $x$, $p$ is some other point satisfying $p \prec q$, and $v$ is a tangent vector at $q$ to the geodesic connecting $p$ and $q$. What this means is that we have a relation
\beq (q, v) \longleftrightarrow (q, \exp_q (- \tau_0 v)) \; , \; \tau_0 >0 \eeq
The minus sign, together with $\tau_0 >0$, is what implies that the ordered pair $(p,q)$ corresponds to the tangent vector at $q$ rather than at $p$. From the point of view of dynamics, what this means is that 
\beq {\rm a) \; If \; p_1 \; and \; p_2 \; are \; far \; away \; from \; each \; other, \; then \; \phi (p_1, q) \; is \; still \; coupled \; to \; \phi (p_2, q)} \nonumber \eeq
But
\beq \rm{ b) \; If \; q_1 \; and \; q_2 \; are \; far \; away \; from \; each \; other, \; then \; \phi (p, q_1) \; is \; \emph{not} \; coupled \; to \; \phi (p,q_2)} \nonumber \eeq
Thats because the ``location" of the segment $(p,q)$ is identified with point $q$ \emph{as opposed to} $p$. The only purpose of $p$ is to simply define a direction of a tangent vector \emph{at $q$}. 

This point is important because, contrary to the other constructions, the distance between $p$ and $q$ is assumed to be \emph{very large} as opposed to very small. The reason for this is that we would like to have some notion of parallel transport of $v$ in order to construct a plane wave. The local edges of discrete set produced by Poisson scatter are highly unlikely to be parallel, even approximately. However, if the distance between $p$ and $q$ is much larger than the distance between $q$ and $q'$, then the vectors $(p,q)$ and $(p,q')$ would, in fact, be approximately parallel. However, approximately parallel isn't enough. After all, if we make a very large number of displacements, each of which is approximately parallel, the small deviations from parallelism would add up. What we want, however, is to be able to speak of two vectors at the two opposite sides of the laboratory that are still approximately parallel to each other. What this means is that the distance between $p$ and $q$ isn't merely large compared to Planck scale but, rather, it is large compared to our laboratory: in particular, it is much larger than the latter. In case of flat space, we may assume that $p$ is the source of an imaginary spherical wave, that later reaches $q$, and the distance between $p$ and $q$ has to be much larger than the scale of laboratory, so that this wave appears to be a plane wave. It is important to note that calling $p$ an ``imaginary source" doesn't make it an actual source: on a contrary, the wave can be produced from other waves, with other imaginary sources, via the coupling in the integral term of Eq \ref{Unconventional}. Furthermore, the ``imaginary spherical wave" analogy doesn't extend to curved space since the solution of the wave equation doesn't match the set of equidistant points. This, however, is not a problem since our actual proposal doesn't involve imaginary waves, we simply made an illustration. 

We can make the distance between $p$ and $q$ to be the one that we want by making sure that when we go from the Lagrangian ${\cal L} (\phi; p, q)$ to action $S(\phi)$, only the pairs $(p,q)$ with the distances that we want are contributing to the sum: 
\beq S (\phi) = \sum_{\{p \prec q \vert \tau_1 < \tau (p,q) < \tau_2 \}} {\cal L} (\phi; p, q) \eeq
where $\tau_1$ and $\tau_2$ are constants. What we just said is that $\tau_1$ has to be large. Now lets ask ourselves how large.  In order to answer this question, we have to first ask ourselves about the distance between $q$ and $q'$. In case of flat Minkowski space, we would like to be able to say that the wave vectors at different parts of the wave are approximately parallel. At the same time, it might not be true about all of them: it is always possible that what appears to be a plane wave is really a spherical wave with a very large radius. In order for us not to notice any difference from a plane wave, however, the radius of a spherical wave has to be much larger than the size of a laboratory. Now, there is an upper bound on the size of the laboratory: namely, it shouldn't be large enough for $\phi^4$-coupling to destroy the plane wave. The effect of $\phi^4$-coupling is $\lambda \phi^4/ (m^2 \phi^2) = \lambda \phi^2/ m^2$. Furthermore, there is a lower bound on the strengths of $\phi$ that \emph{our current equipment} can measure, call it $\phi_{min}$. Thus, the effect of $\phi^4$ coupling that is relevant to what our current equipment can detect is $\lambda \phi_{min}^2/ m^2$. Therefore, the size of the laboratory must be much smaller than the inverse of this: 
\beq L_{Lab} \ll \frac{m^2}{\lambda \phi_{min}^2} \label{Lab} \eeq
In order for our model not to be falsified, we would like to have 
\beq \tau_1 \gg L_{Lab} \label{WantWeaker} \eeq
Since we don't know what $L_{Lab}$ is, we instead postulate a \emph{stronger} condition, 
\beq R = 0 \Longrightarrow \tau_1 \gg \frac{m^2}{\lambda \phi_{min}^2} \label{Stronger} \eeq
The reason we put $R=0$ as one of its premises is that, in case of $R \neq 0$, we have a different constraint on the size of the lab. In particular, it should be small enough for the effects of curvature to be negligible: 
\beq L_{lab} \ll \frac{1}{R} \eeq 
which means that we can weaken Eq \ref{Stronger} as
\beq \tau_1 \gg \min \bigg( \frac{1}{R},  \frac{m^2}{\lambda \phi_{min}^2} \bigg) \label{WeakenedStrong} \eeq
which would still be strong enough to imply Eq \ref{WantWeaker}. At the same time, we would like to be able to define a parallel transport on a distance scale $1/ \sqrt{\alpha}$ (where $e^{- \alpha x^2/2}$ is the Gaussian we will be using to define Laplacian in the integral form in Eq \ref{Laplacian}) If we displace two segments by that distance, we would get a rectangle with area of the order of $\tau_1 /\sqrt{\alpha}$. Since we don't want the effects of the curvature to be felt in that rectangle, we would like to say 
\beq R \ll \frac{\alpha^{1/4}}{\tau_1^{1/2}} \eeq
We can still, however, generalize the model for large $R$ (which might happen due to quantum fluctuations of curvature); we are simply saying that small $R$ are the only ones where the model that we have approximates the continuum counterpart. In case of large $R$ our proposed model would still give well defined answers, they simply won't match anything we know in a continuum. But, if we hypothesize that the discrete model is exact while continuum counterpart is an approximation, then our ``exact" model would work for all values of $R$, while the continuum formula would ``emerge" for the ``special case" where $R$ is small (although, of course, we cheated a little bit since we had that ``special case" in mind while inventing our ``general" formula, plus our ``general" model wasn't tested outside of the ``special case"). 

Let us proceed with the explicit model. If we have two pairs of points, $(p, q)$ and $(p', q')$, such that $p$ is close to $p'$, $q$ is close to $q'$, while $p$ and $q$ (and, therefore, $p'$ and $q'$) are far away from each other, then the geodesic connecting $p$ and $q$ will be almost parallel to the geodesic connecting $p'$ and $q'$. Lets denote those discretized geodesics by $p \prec^* r_1 \prec^* \ldots \prec^* r_{n-1} \prec^* q$ and $p' \prec^* r'_1 \prec^* \ldots \prec^* r'_{n'-1} \prec^* q'$. We can approximate the distance between those two geodesics by emitting the light at a point $r_i$, bouncing it from the point $r'_k$, and receiving it at $r_j$, and then looking at how much time has passed between $r_i$ and $r_j$. In other words, we would like to find $i$, $j$ and $k$ satisfying $r_i \prec^* r'_k \prec^* r_j$ and to record $j-i$. In continuum limit, it doesn't make much difference where we choose to emit the light (or, in other words, which $i$ we pick) since they are parallel. But since there are discrete fluctuations, we might as well minimize their effect by taking average: 
\beq d ((p,q), (p',q'))=  \frac{\xi \sum_{r_i \prec^* r'_k \prec^* r_j} (j-i)}{2 \; \sharp \{r_i \prec^* r'_k \prec^* r_j \}} \eeq
where $\sharp$ stands for number of elements. It can be shown that in $d$-dimensional space 
\beq \Delta f = \lim_{\alpha \rightarrow \infty}  \frac{2 \alpha \int (f (\vec{x}) - f(\vec{0})) e^{- \alpha \vert \vec{x} \vert^2/2} d^dx}{\int e^{- \alpha \vert \vec{x} \vert^2/2} d^d x} \label{Laplacian} \eeq
If we assume that $\Delta f$ on the left hand side is defined in a frame where $v^{\mu} = \delta^{\mu}_0$ and then rotate it back to the general frame, and if we also discretize the right hand side by replacing integral with the sum and removing the limit sign, we obtain 
\beq (g^{\mu \nu} - v^{\mu} v^{\nu}) \frac{\partial^2 \phi}{\partial x^{\mu} \partial x^{\nu}} \approx \frac{2 \alpha  \sum_{(p',q') \in \perp_{\delta} (p,q) } (\phi (p',q')- \phi (p,q)) e^{- \alpha d^2 ((p,q),(p',q'))/2}}{\sum_{(p',q') \in \perp_{\delta} (p,q) }  e^{- \alpha d^2 ((p,q),(p',q'))/2}} \label{ApproximateLaplacian} \eeq
where $\perp_{\delta} (p,q)$ is a set of pairs of points $(p',q')$ that can be thought of as an approximation to the displacement of $(p,q)$ in a direction orthogonal to itself, and $\delta$ is a degree of an approximation; for example, it can be defined as 
\beq \perp_{\delta} (p,q) = \bigg\{(p',q') \bigg\vert p' \prec q', \bigg\{\frac{\tau (p',q')}{\tau (p,q)},  \frac{\tau (p',q)}{\tau (p,q')}\in \bigg((1+ \delta)^{-1}, 1+ \delta \bigg) \bigg\} \bigg\}  \label{OrthogonalTranslation} \eeq
and $\alpha$ is a very large constant. To make sense of the above definition, we say that $(p',q')$ is a displacement of $(p,q)$ in a direction orthogonal to itself if and only if $pqq'p$ form a rectangle. Now, $pqq'p$ form a rectangle if and only if the side $pq$ has the same length as the side $p'q'$ and the diagonal $pq'$ has the same length as the diagonal $p'q$. But, in light of discreteness, the exact equalities is too much to ask for. Instead, we replace them all with approximate statements. We then make those approximate statements precise by replacing $a \approx b$ with $a/b$ is between $(1- \delta)^{-1}$ and $1+ \delta$, for some small $\delta$. Notably, this $\delta$ is finite and, in fact, it is viewed as a physical constant that simply \emph{happened} to be small. We simply don't know what the value of that ``constant" is. In any case, we use Eq \ref{OrthogonalTranslation} and rewrite Eq \ref{Constraint} as 
\beq  \frac{2 \alpha  \sum_{(p',q') \in \perp_{\delta} (p,q) } (\phi (p',q')- \phi (p,q)) e^{- \alpha d^2 ((p,q),(p',q'))/2}}{\sum_{(p',q') \in \perp_{\delta} (p,q) }  e^{- \alpha d^2 ((p,q),(p',q'))/2}} =  - \epsilon (R(p,q)) \phi (p,q) \label{NowExact} \eeq
where $R$ is defined by some discrete method that is beyond the scope of this paper. Notice that, despite the fact that we have an approximation sign in Eq \ref{ApproximateLaplacian}, we have an exact equality in Eq \ref{NowExact}. This is because our philosophy is that physics, as such, is discrete; thus, a continuum is an approximation to the discrete, as opposed to discrete being an approximation to the continuum. Of course, if we are comparing discrete to continuum we would have an approximation sign in both cases, just for different reasons; but if we are comparing one of those two expressions to something physical (such as constraint that we have) we would want to put exact sign in case of discretized expression and an approximate sign in case of continuum one. Thus, we should have had approximation signs in Eq \ref{Unconventional} and \ref{Constraint}, and the only reason we didn't do it is that we were presenting a ``toy model" in order to give the reader an idea of what we are aiming for. Now, the discretization of the derivative in the direction parallel to $v^{\mu}$ is given by 
\beq v^{\mu} \frac{\partial \phi}{\partial x^{\mu}}  \approx \frac{\sum_{(p',q') \in \parallel_{\chi} (p,q)} (\phi (p',q') - \phi (p,q))}{\sharp (\parallel_{\chi} (p,q))} \label{DirectionalDerivative} \eeq
where $\parallel_{\chi} (p,q)$ is a set of vectors $(p',q')$ that can be produced by displacing $(p,q)$ in a direction parallel to itself by a distance $\chi$. In a continuum case, $\parallel_{\chi} (p,q)$ would have exactly one element. But in discrete case such might not be the case: we might have two separate elements of $\parallel_{\chi} (p,q)$ that are very closely aligned next to each other, yet don't exactly match. For that reason, we are simply averaging over all of them. Strictly speaking, we define $\parallel_{\chi} (p,q)$ as 
\beq \parallel_{\chi} (p,q) = \{(p', q') \vert \tau (p,p') = \chi, \tau (p', q) = \tau (p,q) - \chi, \tau (q,q') = \chi, \tau (p',q') = \tau (p,q) \} \eeq 
The fact that $\tau (p,p') + \tau (p', q) = \tau (p,q)$ implies that $p'$ lies on the geodesic segment $pq$. The fact that $\tau (p',q)+ \tau (q,q') = \tau (p',q')$ implies that $q$ lies on the geodesic segment $p'q'$. And, together with $\tau (p,q) = \tau (p',q')$, this implies that $p'q'$ is, in fact a parallel displacement of $pq$. The fact that $p \prec q$ and $p' \prec q'$ implies that said displacement was made in the future direction rather than the past. In any case, if we take the square of Eq \ref{DirectionalDerivative} we obtain
\beq v^{\mu} v^{\nu} \frac{\partial \phi}{\partial x^{\mu}} \frac{\partial \phi}{\partial x^{\nu}} \approx  \bigg( \frac{\sum_{(p',q') \in \parallel_{\chi} (p,q)} (\phi (p',q') - \phi (p,q))}{\sharp (\parallel_{\chi} (p,q))} \bigg)^2 \eeq
Finally, the integral term, corresponding to the interaction, can be discretized as 
\beq  \int dv_1 dv_2 dv_3 \frac{\lambda (v, v_1, v_2, v_3)}{4!} \phi (x, v) \phi (x,v_1) \phi (x,v_2) \phi (x,v_3) \longrightarrow \nonumber \eeq
\beq \longrightarrow \sum_{p_1 \prec q, p_2 \prec q, p_3 \prec q} \frac{\lambda (\tau (p,q), \tau (p_1, q), \tau (p_2, q), \tau (p_3, q) )}{4!} \phi (p,q) \phi (p_1, q) \phi (p_2, q) \phi (p_3,q) \eeq
Substituting our results into Eq \ref{Unconventional}, we obtain the discretization of total Lagrangian:
\beq {\cal L} (\phi; p, q) = \frac{1}{2} \bigg( \frac{\sum_{(p',q') \in \parallel_{\chi} (p,q)} (\phi (p',q') - \phi (p,q))}{\sharp (\parallel_{\chi} (p,q))} \bigg)^2 + \frac{m^2}{2} \phi^2 (p,q) +  \nonumber \eeq
 \beq + \sum_{p_1 \prec q, p_2 \prec q, p_3 \prec q} \frac{\lambda (\tau (p,q), \tau (p_1, q), \tau (p_2, q), \tau (p_3, q) )}{4!} \phi (p,q) \phi (p_1, q) \phi (p_2, q) \phi (p_3,q) \label{Final} \eeq
where, as explained earlier, we put an exact sign because we are assuming that physics is, in fact, discrete. 

\subsection*{Conclusion}

The main thing that was accomplished in this paper is that the role of d'Alembertians was replaced with directional derivatives, which allowed us to regain locality. The key idea is that, instead of assigning the scalar fields to points, we assign it to edges. But the edges to which we assign it are ``very long" and, therefore, its possible for them to be ``almost parallel" to each other -- which wouldn't happen for ``short" edges in case of Poisson distribution of points. This almost-parallelism, in turn, allows us to ``construct" a wavelet by imposing a constraint that the space derivative of $\phi$ assigned to the almost-parallel edges is nearly-zero with respect to any direction almost-perpendicular to the direction of those edges. This is accomplished by making the Laplacian with respect to a hyperplane orthogonal to the direction of the edges to be near-zero. With this constraint in mind, the d'Alembertian in the scalar field Lagrangian can be replaced with the second derivative in the direction parallel to the direction of the edges, which via integration by parts is being replaced by the square of the first derivative with respect to that direction, which can be discretized while avoiding nonlocality. 

The nonlocal effects of the length of the edges are avoided by interpreting the field assigned to an edge as residing at its future-most end. The sole purpose of the ``past" point is to assign the direction of the vector, which is, by default, attached to the ``future" point. By saying that the field is attached to the ``future" point of the edge what we mean is that if we have two edges with nearby future points, they are coupled to each other, but if we have two long edges with nearby past points, then they aren't. The purpose of the past points is to define a direction of the tangent vectors, that are, by default, located at the future points. Now, the interaction between edges with far-away past points and nearby future points is interpreted as an interaction between two tangent vectors, located near each other, but pointing in very different directions. In other words, it amounts to saying that the theory is local in position but not in momentum. Now, the ``non-locality" of momentum is still finite, since there is ultraviolet cutoff, which is implemented through the decay of $\lambda (\cdots)$ in Eq \ref{Final} if $\tau (\cdots)$ gets too large. But the point is that the size of the region of nonlocal interactions in position space is much smaller than the size of the region of nonlocal interactions in momentum space -- which is implemented through the distances between ``future" endpoints of strongly-interacting edges being much smaller than the distances between their ``past" endpoints. Equally importantly, however, both upper bounds have to be finite: we need to have finite value of ultraviolet cutoff in order to avoid divergences. A lot of other causal set proposals, such as \cite{Dambertian1} and \cite{Dambertian2}, are forced to include infinite sums into Lagrangian densities, and thats what we were able to avoid. Of course, however, if we are to go from Lagrangian density to an action we would again have an infinite sum, but that is a different matter. What we are saying is that we have avoided infinite sums at the level of Lagrangian densities. 

Somewhat similar idea was also used in \cite{Bosonic}, but in a different way: in that paper, $v$ was to be aligned with the gradient of $\phi$, while in the current paper the gradient of $\phi$ is to be aligned with $v$. The former leads to non-linear effects (see \cite{Nonlinear}), while the latter doesn't. The source of nonlinearity of \cite{Bosonic} is that when we take path integral over $\phi$, the direction of vector $v$ changes as $\phi$ changes; now, the direction of $v$ represents coupling, thus the coupling change. For any given coupling we have a quadratic form. But since the coupling is a function of $\phi$, the coefficients in a quadratic form are no longer constants but rather are functions of $\phi$, which means that the form isn't quadratic any more which, in turn, makes it impossible to compute the path integral without resorting to numeric simulations. On the other hand, in the paper at hand the couplings are afore-given, which means that the coefficients in quadratic form \emph{are} constants and, therefore, we could compute the path integral analytically if we knew what those constants are. The reason in \cite{Bosonic} we made $v$ to be a function of fields is that in that paper all the scalar information was assigned to points rather than edges (the only thing that was assigned to edges is gauge fields) and since points aren't supposed to have preferred directions (unless we are willing to violate relativity), we had to argue that $v$ isn't a preferred direction assigned to a point but, instead, its the function of field itself and, therefore, its not ``preferred" since it will be integrated over. But the price for it is nonlinearity, as described. In the current paper we also avoided the assignment of preferred frame to a point; but instead of ``shifting blame" to the behavior of the field (which would lead to non-linearity) we ``shifted blame" to the direction of the edge, which avoided the nonlinearity. 

Another paper with a similar concept to the current one is \cite{PhaseSpacetime}. Just like here, the approach of \cite{PhaseSpacetime} is to replace $\phi (x)$ with $\phi (x, v)$. The difference, however, is that in \cite{PhaseSpacetime} the definition of the element of the set was changed from $x$ to $(x,v)$, while in the current paper an element of the set is still $x$, while $(x,v)$ is associated with pair of elements, $(p,q)$. There are arguments in favor of either side. On the one hand, the approach in \cite{PhaseSpacetime} looks more local than here, since in the current paper the two points are separated very far apart. On the other hand, in case of identifying $x$ with an element of the set, one can refer to \cite{Hawking} and \cite{Malament} to argue that there is a one to one correspondence between $\prec$ and $g_{\mu \nu}/ \vert g \vert$; as far as we are aware, there is no analogue of this theorem for $(x,v)$ case. In the worst case scenario, if the version of that theorem for $(x,v)$ is wrong (we don't know one way or the other since we neither proved it nor found a counter example) it would imply some unwanted features such as position displacement being confused with change in velocity. From this perspective, the approach in the current paper is ``safer" one, since it avoids those questions altogether, while the ``nonlocal" nature of $(p,q)$ doesn't have any physical consequences: it is merely our way of identifying a vector. 

There was another paper that, similar to here, modeled $(x,v)$ as $(p,q)$, namely \cite{Edges}, but, unlike the current paper, the distance between $p$ and $q$ in that other paper is small, which makes it look more natural from this perspective. However, in that paper the direction of the gradient of the field was not parallel to $v$. Consequently, we still had to come up with version of d'Alembertian as opposed to directional derivative. Said d'Ambertian was made to be  local: even though the gradient of the field wasn't parallel to $v$, we still ``had" $v$, and ``having" preferred frame allows us to remove nonlocalities one way or the other (see \cite{NonlocalityTheorem}). However, this resulted in Lagrangian density being expressly dependent on both $x$ and $v$, and then we had to remove $v$ dependence by adjusting coefficients in rather clever way to make it cancel. Such adjustment of coefficients might look artificial. In the current paper we don't have to do that: the fact that in the current paper we have made the gradient of $\phi$ parallel to $v$, the unwanted terms that in \cite{Edges} we wanted to cancel, are non-existent on the first place. At the same time, however, one might argue that some of what we do here is artificial in its own way: particularly the fact that in the current paper we are using ``long" edges that are rather unnatural, while in \cite{Edges} the ``short" edges are used, which are a lot more natural. Also, in the current paper we haven't introduced electromagnetic field, while in \cite{Edges} we did. 

Another approach is to simply use non-local d'Alembertian (\cite{Dambertian1} and \cite{Dambertian2}) and argue that the non-local effects will cancel each other out. However, in order for those arguments to go through, one has to implicitly assumed that causal set is bounded. This would imply the preferred frame. After all, the lightcone of a given point would intersect the edges of the causal set, at those points of intersection we can look at the directions of tangent vectors to the edges, and take their average to determine the preferred frame. Essentially, those papers argue that the \emph{effects} of preferred frame would be small, due to cancellation. But there are other discrete theories (see \cite{NonlocalityTheorem}) where the \emph{effects} of non-locality are very small. What puts causal set theory apart is that it \emph{claims} to avoid those \emph{very small} effects that other discrete theories couldn't avoid. The presence of the boundary of a causal set would invalidate such claim. 

One approach that avoids the issues of nonlocality is the one by Johnston (\cite{Johnston1} and \cite{Johnston2}): by focusing on propagators rather than Lagrangians, it avoids the need of d'Alembertian and, at the same time, the propagator is naturally free from non-locality since the only points that contribute are the ones to the future of one of the two points and to the past of the other one. However, it is not clear whether the propagator approach, without the use of Lagrangians, can describe energy momentum tensor and its coupling to gravity, although that is certainly something worth looking into. 

In any case, the bottom line is that I am not saying that the approach given in this paper is necesserely better than the other approaches I just mentioned: as I illustrated there are pluses and minuses on each side. Thats why I think they all need to be on a table, at least for now. And what this particular paper brings to the table is a possibility avoiding nonlinearities (present in \cite{Bosonic} and \cite{Nonlinear}), nonlocalities (present in \cite{Dambertian1} and \cite{Dambertian2}) and artificial adjustment of coefficients (present in \cite{GravityAndMatter} and \cite{Edges}) at the same time and, in contrast to \cite{PhaseSpacetime}, retain the original definition of causal set which would allow the use of \cite{Hawking} and \cite{Malament} to still argue that causal relations are equivalent to geometry. The weakness of the paper at hand is that the edges that are being considered are very long rather than very short, and also that we haven't expanded this approach to gauge field and other non-scalar fields.

\end{document}